\documentclass{acmproc-sp}

\newcommand{\comment}[1]{}
\newcommand{\ket}[1]{\left|#1\right>}
\def\cset{{\Bbb C}}                             
\def\c2n{{{\Bbb C}^{2^n}}}

\def\hX{\hat{{\cal X}}}

\def\mod{\rm \; mod \;}
\newtheorem{lemma}{Lemma}
\newtheorem{theorem}[lemma]{Theorem}

\newtheorem{prop}[lemma]{Proposition}

\def\proof{\noindent {\bf Proof: }}
  
\def\qed{\mbox{\ \ $\Box$}}
\def\rqed{\mbox{ }~\hfill~$\Box$}

\def\cX{\cal X}
\newcommand{\be}{\begin{equation}}
\newcommand{\ee}{\end{equation}}
\newcommand{\bea}{\begin{eqnarray}}
\newcommand{\eea}{\end{eqnarray}}
\newcommand{\bean}{\begin{eqnarray*}}
\newcommand{\eean}{\end{eqnarray*}}
\def\poly{\mbox{poly}}

\def\bbbc{{\mathchoice {\setbox0=\hbox{$\displaystyle\rm C$}\hbox{\hbox
to0pt{\kern0.4\wd0\vrule height0.9\ht0\hss}\box0}}
{\setbox0=\hbox{$\textstyle\rm C$}\hbox{\hbox
to0pt{\kern0.4\wd0\vrule height0.9\ht0\hss}\box0}}
{\setbox0=\hbox{$\scriptstyle\rm C$}\hbox{\hbox
to0pt{\kern0.4\wd0\vrule height0.9\ht0\hss}\box0}}
{\setbox0=\hbox{$\scriptscriptstyle\rm C$}\hbox{\hbox
to0pt{\kern0.4\wd0\vrule height0.9\ht0\hss}\box0}}}}

\newcommand{\V}{{\cal V}}

\newtheorem{claim}{Claim}

\begin{document}

%

\title{
Computing with Highly Mixed States
}
\subtitle{[Extended Abstract]}
%
%

\numberofauthors{3}
\author{
%
\alignauthor Andris Ambainis\thanks{Supported in part by a
U.C. Berkeley Graduate Fellowship
and NSF grant CCR-9800024.
 {\tt ambainis@cs.berkeley.edu}}  \\
\affaddr{Computer Science Division} \\
\affaddr{U. C. Berkeley} \\ 
\affaddr{Berkeley CA 94720}
\alignauthor Leonard J. Schulman\thanks{Supported in part by NSF
       CAREER grant CCR-9876172. {\tt schulman@cc.gatech.edu}}\\
       \affaddr{College of Computing}\\
       \affaddr{Georgia Tech}\\
       \affaddr{Atlanta GA 30332-0280}
\alignauthor Umesh V. Vazirani\thanks{Supported in part by NSF grant
CCR-9800024. {\tt vazirani@cs.berkeley.edu}} \\
\affaddr{Computer Science Division} \\
\affaddr{U. C. Berkeley} \\ 
\affaddr{Berkeley CA 94720}
}
\maketitle

\begin{abstract}
We consider quantum computing in the one-qubit model where 
the starting state of a quantum computer 
consists of $k$ qubits in a pure state 
and $n-k$ qubits in a maximally mixed state.
We ask the following question: is there a general method for simulating
an arbitrary $m$-qubit pure state quantum computation by 
a quantum computation in the $k$-qubit model?
We show that, under certain constraints, this is impossible, unless
$m=O(k+\log n)$.
\end{abstract}

\section{Introduction}

Ideally, a quantum computation is a sequence of local 
unitary transformations applied to a register of qubits
which are initially in the state $\ket{0^n}$; followed
by a measurement.
Initializing the state of the quantum register is the biggest
challenge in NMR quantum computing (which is perhaps the most advanced
technology in terms of the scale of experiments performed to date
\cite{CVZLL98}). The difficulty is that the register is actually
initially in (approximately) the binomial distribution over pure
states $\ket{x}$, in which each qubit is independently in the state
$\ket{0}$ with probability $\frac{1+\epsilon}{2}$; the currently
achievable polarization $\epsilon$ is quite small. There are currently
two ways of implementing quantum computation in this technology. The
first is used in current experiments \cite{GC,CFH}, but does not scale
beyond several qubits --- the output signal decreases exponentially in
the number of qubits in the quantum register.\footnote{
\comment{
* Supported by a U.C. Berkeley Graduate
Fellowship. {\tt ambainis@cs.berkeley.edu}. \\
$\dagger$ Supported in part by NSF
       CAREER grant CCR-9876172. {\tt schulman@cc.gatech.edu}. \\
$\ddag$ Supported by NSF grant CCR
9800024. {\tt vazirani@cs.berkeley.edu}. \\
}
The
exponential decay in signal to noise ratio in any scheme that embeds
virtual pure states on an $n$-qubit quantum computer with one clean
qubit is unavoidable, due to the result \cite{Nielsen}.}
The second does scale,
but is not feasible at the currently achievable values of polarization
in liquid NMR \cite{SV}. An intriguing third possibility was raised in
\cite{KL}. Suppose we start with one qubit in the pure state $\ket{0}$
in tensor product with $n-1$ qubits in a maximally mixed state (i.e.\
in a uniform distribution over basis states $\ket{x}$). Is it possible
to simulate general quantum computation by effecting a sequence of
elementary quantum operations on this register? If the answer were
affirmative, this would yield a procedure that would both scale and be
currently feasible using the scalable initialization procedure to
convert the initial binomial state to a state where the last $n-1$
qubits are maximally mixed and the first bit has high polarization
(the strength of the output signal is now proportional to this
polarization). This is the question we focus on in this paper.

It is easy to see that if all $n$ qubits are in the maximally mixed
state then no computation is possible. This is because applying any
unitary transformation to this mixture leaves it invariant. This
simple argument stands in striking contrast to the difficulty of the
seemingly very similar case, in which just a single qubit is in a pure
state, while all the others are maximally mixed. Since the initial
state of the register is completely specified, the only real input in
this model is the sequence of elementary quantum operations. So, given
a quantum circuit $C$ which we would like to simulate on an input $x$,
we wish to know whether there is a sequence of elementary
quantum operations on the $n$ qubit register, which first prepares a
quantum (mixed) state which encodes $x$, and then simulates $C$ on
it. Of course, we will require that these mixed state encodings of
basis states $x$ be distinguishable by some measurement with
non-trivial probability.

Our main result shows that the above is impossible unless 
$|x| \in O(\log n)$, showing that the simulation is no more 
efficient than an exhaustive classical calculation. The 
technique used to show this uses some information about the 
representations of the symmetric group. The appendix \ref{repthy}
provides some necessary notions from representation theory.

We also show that using a 3-bit register it is possible to 
compute every language in NC1. 
This should give some indication 
of why the impossibility result is so much harder than for the case
when all $n$ qubits are maximally mixed.

\section{NC1}

We begin by showing that in this model, even using a 3-bit register,
we can compute every language in NC1. 

Recall that the initial state of the register is a uniform distribution
over the four $3$-bit strings starting with a $0$. In our simulation
of NC1, all our operations will simply permute basis states.

\begin{prop}
A $3$-bit quantum computer initialized with one clean qubit
can recognize every language in NC1.
\end{prop}

Our simulation is based on Barrington's result that NC1 can
be simulated by a width $5$ permutation branching program \cite{B}. 
The main idea is quite simple: let the $5$ states of 
the permutation branching program be represented by
the states $\ket{000}$ through $\ket{100}$ of the 
$3$-qubit register. Without loss of generality assume that the permutation 
branching program accepts if the permutation it effects is the identity,
and rejects if the permutation it effects is the 
transposition $(000, 100)$. It is easy to simulate the
permutation branching program by a sequence of elementary 
quantum operations. Now if we measure the first
qubit in the register, then in the case that the permutation
branching program accepts -- i.e.\ the permutation effected is
the identity -- then measuring the first qubit in the register
yields a $0$ with probability $1$. On the other hand, if the 
permutation branching program rejects, then measuring the 
first qubit in the register yields a $1$ with probability $1/4$. \rqed

It is illuminating to try to extend this simulation to QNC1. 
First notice that in Barrington's procedure for simulating NC1,
each wire in the NC1 circuit is simulated at some stage in the 
branching program. In the case of a QNC1 circuit, the state of 
a wire is given by a qubit, which is, in general, entangled with
the qubits carried by the other wires in the circuit. Therefore the state 
of this wire cannot be expressed in isolation, and there appears to be
no alternative to creating that entangled state as part of any simulation.
Thus the entire approach breaks down. One way to carry out such a
construction, might be to apply a superposition of operations at each
step: this extends the state space of the quantum computer and
effectively provides many more clean qubits, making the model meaningless.
Moreover 
all proposed implementations of quantum computation
involve a classical, time-varying sequence of operations, applied to a
quantum register. 
Since the control is classical, 
in any oblivious simulation the entangled quantum state
of the simulated circuit must be encoded within the quantum register.

\section{Limit on Computability}

We are given a quantum computer with an $n$ qubit register, with one
bit initialized to $\ket{0}$ and the rest of the $n-1$ qubits in a
maximally mixed state. We would like to simulate an $m$ qubit quantum 
circuit $C$ on input string $x$ using this model. If we wish to do
an oblivious simulation, as sketched in the previous section, we must
encode an arbitrary $m$ qubit state into the uninitialized $n$ qubit
register. To do so, it is sufficient to consider the $2^m$ basis states
of the $m$ wires, and encode them as distinguishable states of the
uninitialized $n$ qubit register (for this to be an efficient encoding,
we should have $n \in O(poly(m))$). The states must be distinguishable
in the following  
sense: since we can prepare several copies of any state by repeating
the simulation, we only require that there be a sequence of measurements
on $O(poly(n))$ many copies of the state, that (with high probability)
uniquely identify the state.
Indeed, it is possible to do this with $n = m$, as follows: 
take the subspaces, spanned by the basis vectors in the
sets $A_b = \{ x \in \{0,1\}^n: x \cdot b
= 0 \mod 2 \}$ for $b \in \{0,1\}^n$. 

However, to perform an oblivious simulation, the encoding must satisfy
another property -- permutability. The quantum circuit $C$ might carry
out any unitary operation on its quantum state, and in particular an
arbitrary permutation on its classical states. Again it is not hard to
demonstrate an efficient encoding that satisfies this permutability
condition, without distinguishability:
take the subspaces
spanned by the basis vectors in the sets $A_b= \{ x=(x_1...x_n) \in
\{0,1\}^n: x_1=0 \mbox{ or } (x_2...x_n)=b \}$  for $b \in \{0,1\}^{n-1}$.

However, it is not possible to construct an efficient encoding that 
satisfies both conditions simultaneously. This is the content of the 
following theorem.

Let $M=2^m$ be the total number of basis states of the ideal quantum
computer which is being simulated. Note that each $X \in \cX$ encoding
one of these, is a subspace of dimension $2^{n-1}$ within the Hilbert
space $\cset^{2^n}$ of the computer. If the computer has $k$ clean
qubits, then $X$ is of dimension $2^{n-k}$.

\begin{theorem}
\label{main-theorem}
Suppose that computations on $m$ qubits can be obliviously simulated in
an $n$-qubit, $k$-clean-qubit computer in our model, and that $\dim(X
\cap Y)/\dim(X) < 1-{1 \over \poly(m)} $ for every
pair of input encodings $X,Y \in \cX$. Then $m \leq (2k+\log n)(1+o(1))$.
\end{theorem}

This incidentally implies that the computation of an $n$-qubit,
$k$-clean-qubit computer can be simulated by a classical computer with
a $\poly(n2^{k})$ computational overhead.

It may be illuminating to consider a
simpler, classical analogue of our problem. A classical
circuit (taking inputs in $\{0,1\}^n$) composed of reversible gates
executes a permutation of $\{0,1\}^n$. The analogous problem (just
considering the case $k=1$) is that we can only represent inputs as
uniform probability distributions over a set of half the elements of
$\{0,1\}^n$. (In the quantum case this 
corresponds to axis-parallel subspaces of dimension $2^{n-1}$.) The
question is, what is the largest number of such subsets (probability
distributions) which such a circuit can permute at will. It is also
essential that the probability distributions be readily
distinguishable by sampling, in other words the subsets must have
small intersection. It is possible (though we omit it in this extended
abstract) to provide a strictly combinatorial argument expressing the
fact that this task is impossible for more than $\poly(n)$ subsets,
because of the tension between the two requirements (permutability and
distinguishability). The large size of the subsets means that we have
far more constraints than we have degrees of freedom. The
combinatorial argument shows that if the requirement of full
permutability is imposed, and we have a superpolynomial (in $n$)
number of subsets, then the symmetric difference of every two sets
must be a vanishing fraction of the size of the
sets. The two types of sets $\{A_b\}$ described above, however,
separately achieve distinguishability and permutability.

In the quantum case we have arbitrary subspaces in place of
``subsets'' (or correspondingly axis-parallel subspaces).
And the circuit of course can perform not just permutations of the basis,
but general unitary operations. In sharp contrast with the classical
case, two subspaces of half the dimensionality of the space typically
will not intersect. Nevertheless, the large dimension of the subspaces
imposes strict constraints on an operator which must permute them; the
difficulty is in formulating the incompatibility of these requirements
when the number of subspaces is large and the subspaces are required
to be very distinct.

\proof 

By assumption, there are unitary operators (each corresponding
to some sequence of steps in the computer) permuting $\cX$ in all
ways. Let $f_{\pi}$ be the unitary operator corresponding to
a permutation $\pi\in S_M$.
If we have $f_{\pi\sigma}=f_{\pi}f_{\sigma}$ 
for all $\pi$ and $\sigma\in S_M$, then
the operators $f_{\pi}$ form a representation of $S_M$ and
we can apply the representation theory of the symmetric group.

Actually, the situation is slightly more complicated.

Let $U$ be the unitary group on $\cset^{2^n}$. Let $H$ be the
subgroup of $U$ acting on $\cX$, i.e.\ carrying any $X \in \cX$ to
some $Y \in \cX$. Let $G$ be the subgroup of $H$ that fixes all of
$\cX$; thus $G$ is normal in $H$.

It is apparent that $H/G \cong S_M$, but although this means that $H$ can
permute the subspaces $\cX$ in arbitrary ways, it is different from
saying that there is a subgroup of $H$ isomorphic to $S_M$ (or in
other words that we can pick elements of $H$ so as to have these
operators compose properly).

\section{Proof of Theorem 2: the simple case}
\label{Simple}

First, we show how to prove Theorem \ref{main-theorem}
if we can select transformations $f_{\pi}$ so that
they form a representation ($f_{\pi} f_{\sigma}=f_{\pi\sigma}$).
The more general case will be handled in the next section.
Appendix (section \ref{repthy}) explains the notions of
representation theory used in this and the next section.

We show that every pair $X,Y \in \hX$ have a substantial
intersection. Consider the decomposition of $\cset^{2^n}$
into irreducible representations $\rho_1 \oplus ... \oplus
\rho_k$. 
Let $N=2^n$.

\begin{lemma}
Either the first row or the first column of the Young diagram of each
$\rho_i$ is of length more than $M-c n$.
\end{lemma}
\proof

This results follows from a theorem by Rasala\cite{Rasala}:

\begin{theorem}\cite[pp.151-152]{Rasala}
\label{RT}
\begin{enumerate}
\item
Let $A\leq M/2$ and $\rho$ be an irreducible representation
of $S_M$ such that the first row of the Young diagram of $\rho$ is
of length exactly $M-A$. Then,
\[ \dim\rho \geq \varphi_A(M) \]
where $\varphi_A(M)={M \choose A}-{M \choose A-1}=
\frac{M-2A-1}{M-A-1} {M\choose A}$ is the dimension of 
the irreducible representation corresponding to the partition 
$(M-A, A)$.
\item
If $\rho$ is an irreducible representation with both the first
row and the first column of length at most $M/2$, then
\[ \dim\rho\geq \varphi_{\lfloor M/2\rfloor}(M) .\] 
\end{enumerate}
\end{theorem}

This theorem means that any representation with the first
row of the Young diagram having length at most $M-k$, $k\leq M/2$
has dimension at least 
\[ \min_{B: A\leq B\leq M/2} \varphi_B(M) .\]
Simple algebra shows that this expression is minimized by $B=A$ if
$A\leq M/2-c\sqrt{M}$ for some constant $c$ and $B=\lfloor M/2\rfloor$
if $A> M/2-c\sqrt{M}$.\footnote{In the second case, the lowest-dimensional
representation actually has the first row less than $M-A$.
It is quite surprsing because, in most cases, removing a square from
the first row of a Young diagram and adding a square somewhere else
increases the dimension.}

To deduce our lemma, assume that the Young diagram of 
an irreducible representation of $S_M$ 
has both first row and column
of length at most $M-A$.
We show that $N\geq 2^A$.
Consider two cases:

{\em Case 1:} $A\geq M/2-c\sqrt{M}$.

Notice that $A\leq M-\sqrt{M}$ because otherwise the Young diagram 
would fit into a square with a side less than $\sqrt{M}$ and
area less than $M$. Theorem \ref{RT} implies that 
\[ \dim \rho\geq \varphi_{\lfloor M/2\rfloor}(M) = 
\Omega\left( \frac{2^{M}}{M\sqrt{M}} \right) = 
2^{M-\frac{3\log M}{2}-O(1)} > 2^A.\]

{\em Case 2:} $A\leq M/2-c\sqrt{M}$. 
Then, 
\[ \dim\rho\geq \varphi_A(M) = \frac{M-2A+1}{M-A+1} {M \choose A} \geq
\frac{1}{M} \left( \frac{M}{A} \right)^A \]
\[ = \frac{1}{M} \left(\frac{M}{2A}\right)^A 2^A \geq 2^A .\]
\rqed

Another lower bound on the
longest row or column of a low dimension representation 
(for a different range of parameters) was given 
by Mischenko\cite{M}. 
 
Next consider the stabilizer of $X$ in $S_M$, which is isomorphic to
$S_{M-1}$, and which we will denote $S_{M-1}^X$. $X$ decomposes into
irreducible representations $V_1,...V_\ell$ of $S_{M-1}^X$. 
$V_1$ is carried by $S_M$ into
each $Y \in \hX$, and all these copies of $V_1$ are contained within
some irreducible $W$ of $S_M$ in $\cset^{2^n}$.

By the previous lemma, the Young diagram of $W$ has a long first
column or row. We use the following fact from the representation
theory of the symmetric group: 
when we restrict an irreducible representation0
$\rho_\lambda$ of $S_M$ of shape
$\lambda$ to a subgroup $S_{M-1} \subseteq S_M$, it decomposes into
irreducibles of $S_{M-1}$ in the following way:
\[ \rho = \bigoplus_{\lambda^{-}} \rho_{\lambda^{-}} \]
where ${\lambda^{-}}$ ranges over all shapes of size $M-1$ that can be
obtained by deleting an ``inside corner'' from $\lambda$. (An inside
corner is simply a point of the shape whose deletion leaves a legal
shape.)

We now use:
\begin{lemma} \label{shape-implic}
Suppose the shape $\lambda$ has a first row (column) of length
$|\lambda|-\ell$ for $\ell<|\lambda|/2$.  Let $\lambda_1$ denote the
``$\lambda^{-}$'' obtained by deleting the last element of the first
row (column). Then $\dim(\rho_{\lambda_1}) \geq {|\lambda|-2\ell \over
|\lambda|} \dim(\rho_{\lambda})$.
\end{lemma}
\proof 
Consider the ratio
${\dim \rho_{\lambda_1} \over \dim \rho_{\lambda}} = {1 \over
|\lambda|} {\prod_{x \in \lambda} |x| \over \prod_{x \in \lambda_1}
|x|}$. In the last ratio, points $x$ outside of the last row (column)
appear identically in the numerator and denominator. Moreover for each
$x$ in the first row (column) in the numerator other than the very
last point (which contributes a factor of $1$ in the numerator and is
absent in the denominator), the ratio between its contributions in the
numerator and denominator is ${|x| \over |x|-1}$. Just examining the
$|\lambda|-2\ell-1$ points of the first row furthest from the
upper-left corner (and excepting the last point), we obtain a lower
bound on these contributions of $\prod_{i=1}^{|\lambda|-2\ell-1} {i+1
\over i}=|\lambda|-2\ell$. Overall therefore ${\dim \rho_{\lambda_1}
\over \dim \rho_{\lambda}} \geq {|\lambda|-2\ell \over |\lambda|}$.
\rqed

\begin{lemma}
\label{bound-difference}
Let $f_{\pi}$ be an $N$-dimensional 
representation of $S_M$ that acts as a permutation
representation on a collection of $M$ subspaces $\hX$.
Then, for any $X, Y\in \hX$, 
\[ \dim X- \dim X\cap Y \leq \frac{2cn}{M} N .\]
\end{lemma}
(We write $n=\lg N$ and $m = \lg M$.)

\proof
We decompose $f_{\pi}$ into irreducible representations.
Let $W$ be one of these irreducible representations.
We show that $\dim X \cap W-\dim X\cap Y\cap W\leq \frac{2c n}{M} \dim W$. 

We look at $W$ as a representation of $S_{M-1}^X$. 
Let $V$ be the highest dimensional irreducible representation of
$S_{M-1}^X$ within $W$.
If $M<2cn$ then the assertion is trivial.
Otherwise $M-c n>M/2$ and the
hypothesis of lemma \ref{shape-implic} is satisfied, implying that 
$\dim V \geq {M - 2c n \over M} \dim W$.
Consider two cases:

\noindent
{\bf Case 1:}
$V\subseteq X$.

Take $\pi\in S_M$ such that $Y=f_\pi(X)$.
Then, $Y\cap W = f_{\pi}(X\cap W)$.
Therefore, $\dim (Y\cap W)=\dim (X \cap W) \geq \dim V\geq 
{M - 2c n \over M} \dim W$ and
\[ \dim X\cap W-\dim X\cap Y\cap W \leq 
{2c n \over M} \dim W .\]

\noindent
{\bf Case 2:}
$V\not\subseteq X$.

Then, 
$V\cap X=0$ 
because $V\cap X$ is invariant under $S^X_{M-1}$
and $V$ is irreducible. $V\cap X=0$ implies
\[ \dim V +\dim X\cap W\leq \dim W .\]
Together with $\dim V\geq \frac{M-2cn}{M} \dim W$, this implies
\[ \dim X\cap W-\dim X\cap Y\cap W\leq \dim X\cap W\leq \frac{2cn}{M} \dim W.\]

The lemma follows 
by summation over all irreducible $W$.
\rqed

If $f_{\pi}$ form a representation, Lemma \ref{bound-difference} 
almost immediately implies Theorem \ref{main-theorem}.
Namely, we have 
\[ \frac{\dim X\cap Y}{\dim X}
= \frac{\dim X-(\dim X-\dim X\cap Y)}{\dim X}
\] \[
\geq \frac{2^{n-k} - \frac{2cn}{M} 2^n}{2^{n-k}} 
= 1-\frac{2^{k+1}cn}{M} .\]
If this is at most $1-\frac{1}{\poly(m)}$, then 
$m \leq (k+\log n)(1+o(1))$.
\rqed

\section{Proof of Theorem 2: the difficult case}
\label{subgp}

\subsection{Proof outline}

Next, we deal with the case when $f_{\pi} f_{\sigma}\neq f_{\pi\sigma}$
for some $\pi$ and $\sigma\in S_M$.
Let $G$ be the group of transformations
that map every subspace $X\in\hX$ to itself. 
Then, $f_{\pi}f_{\sigma}f^{-1}_{\pi\sigma}$ is an element of $G$
for any $\pi, \sigma\in S_M$.
We would like to modify $f$ so that this element becomes identity
for all $\pi$ and $\sigma\in S_M$.
Then, $f_{\pi}f_{\sigma}=f_{\pi\sigma}$, i.e.,
$f_{\pi}$ would form a representation of $S_M$ and 
we would be able to analyse this representation similarly to the previous
section. 

To achieve this,
we look at $\bbbc^{2^n}$ as a representation of $G$ and
express $\bbbc^{2^n}$ as $V_1\oplus V_2 \ldots \oplus V_k$,
with $V_i$ corresponding to different types of irreducible
representations of $G$.

Then, we compose each $f_{\pi}$ with an appropriate $g_{\pi}\in G$.
The resulting transformation $f'_{\pi}=g_{\pi}f_{\pi}$ still
implements the same permutation $\pi$ of $\hX$
because $g_{\pi}$ maps every $X\in\hX$ to itself. 
We can choose the transformations $g_{\pi}$ so that,
on every $V_i$, $f'_{\pi}f'_{\sigma}$ is the same as 
$f'_{\pi\sigma}$ up to a phase 
($f'_{\pi\sigma}=c_{\pi,\sigma, i}f'_{\pi}f'_{\sigma}$ for
some unit 
$c_{\pi,\sigma, i}\in\bbbc$).

The next step is eliminating the phase factors $c_{\pi, \sigma, i}$.
This is done by considering a larger space 
$V_1\otimes V^{*}_1+\ldots+V_k\otimes V^{*}_k$ and
transformations $f''_{\pi}=f'_{\pi}\otimes (f'_{\pi})^*$
on this larger space. 
Then, the phase factors $c_{\pi, \sigma, i}$ (from $f'$) and
$c^*_{\pi, \sigma, i}$ (from $(f')^{*}$) cancel out
and we get $f''_{\pi\sigma}=c_{\pi, \sigma, i}c^*_{\pi, \sigma, i}
f''_{\pi}f''_{\sigma}=f''_{\pi}f''_{\sigma}$.
Thus, $f''_{\pi}$ form a representation of $S_M$ on
the linear space $V_1\otimes V^{*}_1+\ldots+V_k\otimes V^{*}_k$.
This representation can be analysed similarly to section \ref{Simple},
obtaining lower bounds on intersections of invariant subspaces.

\subsection{Representation up to phases $c_{\pi, \sigma, i}$}

Let $G$ be the group of unitary transformations that fix 
every one of the subspaces $X\in\hX$.

Then, $\cset^{2^n}$ is a representation of $G$ and all $h\in\hX$
are invariant subspaces. (They are fixed by every element of
$G$ according to the definition of $G$.)
These invariant subspaces decompose into irreducible invariant 
subspaces.

Consider all the irreducible invariant subspaces of $\cset^{2^n}$.
Split them into equivalence classes consisting of isomorphic 
irreducible subspaces.
Let $E_1, \ldots, E_k$ be these equivalence classes.
Let $V_1$ be the subspace of $\cset^{2^n}$ spanned by all the irreducible
subspaces
in $E_1$ (i.e., the subspace spanned by all the vectors belonging
to at least one subspace in $E_1$). Let $V_2$, $\ldots$, $V_k$ be
defined similarly.

\begin{claim}
If $i, j\in\{1, \ldots, k\}$ and $i\neq j$, then $V_i\perp V_j$.
\end{claim}

Therefore, $\cset^{2^n}=V_1\oplus V_2 \oplus \ldots \oplus V_k$.
Next, we show that transformations $f_{\pi}$ map
each $V_i$ to some (possibly different) $V_{i'}$.

\begin{claim}
\label{Iso}
Let $V$ be an invariant subspace.
Then, $f_{\pi}(V)$ is invariant as well. If $V$ is irreducible,
$f_{\pi}(V)$ is irreducible. Moreover, if $V$ and $V'$ are two 
isomorphic irreducible subspaces, $f_{\pi}(V)$ and $f_{\pi}(V')$
are isomorphic as well.
\end{claim}

\proof
The map $g\rightarrow f_{\pi} g f^{-1}_{\pi}$ is an automorphism
of $G$. If $V$ is invariant under the action of $g$, $f_{\pi}(V)$
is invariant under the action of $f_{\pi} g f^{-1}_{\pi}$.
Therefore, if $V$ is invariant under $G$, so is $f_{\pi}(V)$.

If $f_{\pi}(V)$ is not irreducible, it decomposes into two or more invariant
subspaces: $f_{\pi}(V)=W_1\oplus W_2$. Then, $f_{\pi}^{-1}(W_1)$ is 
invariant as well, implying that $V$ is not irreducible.

Finally, let $h:V\rightarrow V'$ be a $G$-isomorphism of $V$ and $V'$
(an isomorphism that commutes with the action of $G$). 
Let $h':f_{\pi}(V)\rightarrow f_{\pi}(V')$ be defined by
$h'=f_{\pi} h f^{-1}_{\pi}$. Then, for any $g=f_{\pi} g' f^{-1}_{\pi}$,
we have
\[ h'g=(f_{\pi} h f^{-1}_{\pi}) (f_{\pi} g' f^{-1}_{\pi}) =
 f_{\pi} hg' f^{-1}_{\pi} = f_{\pi} g'h f^{-1}_{\pi} = gh' \]
and every $g\in G$ can be expressed in the form $f_{\pi} g' f^{-1}_{\pi}$.
Therefore, $h'$ is a $G$-isomorphism of $f_{\pi}(V)$ and $f_{\pi}(V')$.
\rqed

\noindent {\bf Remark.} 
$V$ does not have to be isomorphic to $f_{\pi}(V)$ as a 
representation of $G$. 
$f_{\pi}$ establishes the isomorphism of $g$ on 
$V$ with $f_{\pi} g f^{-1}_{\pi}$ on $f_{\pi}(V)$, but 
$f_{\pi} g f^{-1}_{\pi}$ does not have to equal $g$ 
on $f_{\pi}(V)$.

\begin{claim}
\label{VSub}
For every $i\in\{1, \ldots, k\}$ there is an $i'$ such that 
$f_{\pi}(V_i)=V_{i'}$.
\end{claim}

\proof
By Claim \ref{Iso}, every two isomorphic irreducible subspaces
get mapped to isomorphic irreducible subspaces.
Therefore, all subspaces in $E_i$ get mapped to subspaces
in the same $E_{i'}$ and $f_{\pi}(V_i)\subseteq V_{i'}$. 
Similar reasoning applied to $f_\pi^{-1}$ implies
$f^{-1}_{\pi}(V_{i'})\subseteq V_i$.
\rqed

For each $i\in\{1, \ldots, k\}$, $V_i$ is the direct sum of some
number of isomorphic irreducible subspaces: $V_i=V_{i1}\oplus V_{i2}\oplus
\ldots V_{ij_i}$. We fix $G$-isomorphisms $h_{ijj'}$ 
between $V_{ij}$ and $V_{ij'}$ so that $h_{ij'j''}h_{ijj'}=h_{ijj''}$.
(By Schur's lemma, each of these isomorphisms is unique up to a
multiplicative constant. The isomorphisms can be made to compose 
properly by adjusting these constants. Note of course that $h_{ijj}$
is the identity.)

\begin{claim}
\label{Form}
$W\subseteq V_i$ is an irreducible invariant subspace if and only if
\[ W=\{a_j x +a_{j+1} h_{ij (j+1)}(x)+\ldots+a_{j_i} h_{ijj_i}(x) | x\in V_{ij}\} \]
for some $j\in\{1, \ldots, j_i\}$ and $a_j, \ldots, a_{j_i}\in \bbbc$.
\end{claim}

\proof
``If'' part:

Invariance: 
\[ g( \sum_{\ell=j}^{j_i} a_{\ell} h_{ij\ell}(x) )
= \sum_{\ell=j}^{j_i} a_{\ell} g(h_{ij\ell}(x)) 
=\sum_{\ell=j}^{j_i} a_{\ell} h_{ij \ell}(g(x)) \]
because each $h_{ij\ell}$ is a $G$-isomorphism.

$W$ is irreducible because, if 
$W_1\subset W$ and $W_1$ is invariant, then 
\[\{ x | a_j x +a_{j+1} h_{ij(j+1)}(x)+\ldots+a_{j_i} h_{ijj_i}(x) \in
W_1 \} \] 
is an invariant subspace of $V_{ij}$; but
$V_{ij}$ is irreducible and $\dim(W)=\dim(V_{ij})$.

``Only if'' part:

Let $W$ be an irreducible invariant subspace of $V_i$.
Let $x'\in W$. Then, we can write $x'$ as $x'_1+\ldots+x'_{j_i}$,
$x'_1\in V_{i1}$, $\ldots$, $x'_{j_i}\in V_{ij_i}$.
If $x' \neq x'' \in W$, then for any index $j$, $x'_j\neq x''_j$ or
$x'_j=x''_j=0$. (Otherwise, $W\cap \bigoplus_{\ell \neq j} V_{i\ell}$
is a nontrivial subspace
of $W$. It is invariant because $V_i$ and $V_{ij}$ are invariant. 
Contradiction with the irreducibility of $W$.)

Let $j$ be the smallest index for which there is an $x'\in W$ with
$x'_j\neq 0$. Then, for every $x\in V_{ij}$, there is an $x'\in W$
with $x'_j=x$. 
(For, if $A$ and $B$ are invariant subspaces of a unitary representation,
the projection of $A$ onto $B$ is invariant. Apply this with
$A=W$ and $B=V_{ij}$, then use the irreducibility of $V_{ij}$.)

The above considerations allow us to define the mapping $h_{j j'}:
V_{ij}\rightarrow V_{ij'}$  by $h_{jj'}(x'_j)=x'_{j'}$. By the
definition, $h_{j j'}(g(x'_j))=h_{j j'}((g(x'))_j)=(g(x'))_{j'}=
g(x'_{j'})=g(h_{j j'}(x'_j)$, so $h_{j j'}$ is a $G$-isomorphism.
By Schur's lemma, this implies that $h_{jj'}=a_{j'} h_{ijj'}$ for
some $a_{j'}\in\bbbc$. 
\rqed

In general, if $W$ is any irreducible invariant subspace of $\cset^{2^n}$,
then $W$ must be in the form described by claim \ref{Form}
for some $i$. ($W$ belongs to some equivalence class $E_i$ and
therefore is contained in the corresponding $V_i$.)

For each $V_{i1}$ ($i\in\{1,\ldots,k\}$), we fix an orthonormal basis
$v_{i1}, \ldots, v_{it}$.
This also fixes a related basis $h_{i1j}(v_{i1})$, 
$\ldots$, $h_{i1j}(v_{it})$ for each $V_{ij}$. 
Moreover, we also get a similar basis
\[ \sum_{\ell=j}^{j_i} a_\ell h_{i1\ell}(v_{i1}),\ldots,
    \sum_{\ell=j}^{j_i} a_\ell h_{i1\ell}(v_{it})\]
for every invariant irreducible $W\subseteq V_i$ because
any such $W$ can be written in the form given by claim \ref{Form}.
We call these bases {\em designated}.

This designated basis is exactly the basis for $W$ that can be obtained
by applying the isomorphism between $V_{i1}$ and $W$ 
to the basis for $V_{i1}$. 
Moreover, if $W, W'$ are two isomorphic irreducible subspaces,
the designated basis for $W$ is mapped to the designated basis for $W'$
by the isomorphism between $W$ and $W'$.
 
We are going to impose the following condition on $f'_{\pi}$:

{\bf Condition.}
Let $W$ be an irreducible representation of $G$ and $w_1, \ldots, w_l$ be
the designated basis of $W$. 
Let $w'_1, \ldots, w'_l$ be the designated basis 
of $f'_{\pi}(W)$. 
Then, there exists $c\in \cset$, $|c|=1$ such that
$f'_{\pi}(w_1)=c w'_1$, $\ldots$, 
$f'_{\pi}(w_l)=c w'_l$.

Next, we show that this condition suffices to guarantee 
$f'_{\pi}f'_{\sigma}=c_{\pi, \sigma, i} f'_{\pi\sigma}$ on every $V_i$ 
and that any $f_{\pi}$ that 
permutes $X\in\hX$ without satisfying this condition
can be transformed into $f'_{\pi}$ that satisfies the condition and
still permutes the subspaces in the same way.

First, we show that it is enough to ensure that the designated basis
of $V_{i1}$ is mapped correctly for every $i\in\{1,\ldots, k\}$.

\begin{claim}
\label{Design}
Assume that the condition is true for $W=V_{i1}$.
Then, it is also true for any irreducible $W\subseteq V_i$. 
\end{claim}

\proof
Let $h$ be the isomorphism between $V_{i1}$, $W$.
Note that $h$ maps the designated basis of $V_{i1}$ to
the designated basis of $W$.

Then (by claim \ref{Iso}) 
 $f'_{\pi}h (f'_{\pi})^{-1}$ is an isomorphism between 
$f'_{\pi}(V_{i1})$ and $f'_{\pi}(W)$.
We know that there is an isomorphism between 
these two irreducibles that maps the designated basis
of one of them to the designated basis of the other.
By Schur's lemma, any two isomorphisms of irreducible 
subspaces can differ only by a multiplicative constant $c$.
The unitarity of $f'_{\pi}h (f'_{\pi})^{-1}$ implies that
$|c|=1$.

Therefore, $f'_{\pi}h (f'_{\pi})^{-1}$ maps the designated 
basis of $f'_{\pi}(V_{i1})$ to $c$ times 
the designated basis
of $f'_{\pi}h (f'_{\pi})^{-1}(f'_{\pi}(V_{i1}))=
f'_{\pi}h (V_{i1})=f'_{\pi}(W)$.
We know that $(f'_{\pi})^{-1}$ maps the designated basis of
$f'_{\pi}(V_{i1})$ to the designated basis of $V_{i1}$ and
that $h$ maps the designated basis of $V_{i1}$ to the designated
basis of $W$. 
This implies that $f'_{\pi}$ maps the designated basis
of $W$ to $c$ times the designated basis of $f'_{\pi}(W)$.
\rqed

\medskip
Next, we show how to transform $f_{\pi}$ into $f'_{\pi}$
that performs the same permutation $\pi$ of $\hX$
and maps the designated basis of every $V_{i1}$ as required.

Let $W_1$, $\ldots$, $W_k$ be $f_{\pi}(V_{11})$, $\ldots$, $f_{\pi}(V_{k1})$.
Each of $W_i$ lies within one of $V_1, \ldots, V_k$.
Denote this subspace $V_{i'}$.
Then, for $i\neq j$, $V_{i'}\neq V_{j'}$.
For each $i\in\{1, \ldots, k\}$, we define a unitary transformation 
$g_{\pi, i}$ on $V_i$ such that $g_{\pi, i'} f_{\pi}$
maps the designated basis of $V_{i1}$ to the designated basis of
$f_{\pi}(V_{i1})$.

By Claim \ref{Form}, the irreducible subspace $W_i=f_{\pi}(V_{i1})$
is just 
\[ \{ a_{i'j} x+a_{i'(j+1)} h_{i'j(j+1)}(x)+\ldots + a_{i'j_{i'}}
h_{i'jj_{i'}}(x) | x\in V_{i'j} \} \]
for some $j$. Moreover, the mapping that maps each $v\in W_i$ to
its $V_{i'j}$-component is an isomorphism of $W_i$ and $V_{i'j}$
w.r.t. $G$ (similarly to proof of Claim \ref{Form}).

Let $v_1, \ldots, v_l$ be the designated basis of $V_{i1}$,
$v'_1, \ldots, v'_l$ be $f_{\pi}(v_1), \ldots, f_{\pi}(v_l)$
and $v''_1$, $\ldots$, $v''_{l}$ be the $V_{i'j}$ components of
$v'_1$, $\ldots$, $v'_{l}$.

Let $w_1$, $\ldots$, $w_l$ be the designated basis of $V_{i'j}$ and
$g_{\pi,i'j}$ be the unitary transformation on $V_{i'j}$
that maps $v''_1, \ldots, v''_l$ to $w_1$, $\ldots$, $w_l$.
We define a unitary transformation $g_{\pi,i'j'}$ (for every $j'\neq j$) 
on $V_{i'j'}$ to be $h_{i'jj'`}g_{\pi,i'j}h^{-1}_{i'jj'}$.
Finally, we take the transformation $g_{\pi,i'}$ of $V_{i'}$ that
is equal to $g_{\pi,i'j}$ on each $V_{i'j}$. 
Then, $g_{\pi,i'}$ maps 
\[ v'_1= a_{i'j} v''_1+a_{i'(j+1)} h_{i'j(j+1)}(v''_1)+\ldots + a_{i'j_{i'}}
h_{i'jj_{i'}}(v''_1) \]
to 
\[ a_{i'j} g_{\pi,i'j}(v''_1)+a_{i'(j+1)} g_{\pi,i'(j+1)}
h_{i'j(j+1)}(v''_1) +\ldots = \]
\[ a_{i'j} g_{\pi,i'j}(v''_1)+ a_{i'(j+1)}
  h_{i'j(j+1)}g_{\pi,i'j}(v''_1)+\ldots\] 
\[ =a_{i'j} w_1 + a_{i'(j+1)} h_{i'j(j+1)}(w_1)+ \ldots \]
which is exactly the first vector of the designated basis for $W_i$.
The same is true for $v'_2$, $\ldots$, $v'_l$, implying
that $g_{\pi,i'}f_\pi$ maps the designated basis of $V_{i1}$ to the
designated basis of $W_i$.

Now, we take $g_{\pi}$ that is equal to $g_{\pi,i}$ on each $V_i$
and take $f'_{\pi}=g_{\pi} f_{\pi}$.

\begin{claim}
$g_{\pi}$ preserves all $X\in\hX$.
\end{claim}

\proof
By definition, the restriction $g_{\pi}\vert_{V_i}$ is equal to
$g_{\pi,i}$, and $g_{\pi,i}$ clearly preserves $V_{i1}, \ldots, V_{ij_i}$.
Moreover, $g_{\pi,i}$ (and, hence, $g_{\pi}$)
preserves any irreducible subspace 
$W\subseteq V_i$ because any such subspace is in the form
of claim \ref{Form}. 

Every $X\in\hX$ is invariant under $G$. Therefore, it decomposes into
a direct sum of irreducible subspaces. Each of these subspaces
is in one of the classes $E_1$, $\ldots$, $E_k$ and, therefore,
lies in one of $V_1$, $\ldots$, $V_k$.
This means that it is preserved by $g_{\pi}$.
Therefore, $X$ which is a direct sum of such irreducible
subspaces is preserved by $g_{\pi}$ as well.
\rqed

Hence, $f'_{\pi}=g_{\pi} f_{\pi}$ realizes the same permutation 
$\pi$ of $X\in\hX$ as $f_{\pi}$.

\begin{claim}
On every $V_i$,
$f'_{\pi}f'_{\sigma}=c_{\pi, \sigma, i}f'_{\pi\sigma}$
for some $c_{\pi, \sigma, i}\in\cset$.
\end{claim}

\proof
This is equivalent to showing that
$(f'_{\pi\sigma})^{-1} f'_{\pi} f'_{\sigma}$
is equal to $c_{\pi, \sigma, i}$ times the identity. 
To show that, notice that $(f'_{\pi\sigma})^{-1} f'_{\pi} f'_{\sigma}$
maps every subspace $X\in\hX$ to itself because $(f'_{\pi\sigma})^{-1}$
performs the inverse of the permutation $\pi\sigma$ on
$\hX$.
Therefore, $(f'_{\pi\sigma})^{-1} f'_{\pi} f'_{\sigma}\in G$.
This means that $V_{ij}$ are all preserved by 
$(f'_{\pi\sigma})^{-1} f'_{\pi} f'_{\sigma}$.

Moreover, $f'_{\sigma}$, $f'_{\pi}$ and $f^{-1}_{\pi\sigma}$
all map the designated bases to $c$-times designated
bases (Claim \ref{Design}).
\comment{the designated basis of $V_{ij}$ is mapped to
the designated basis of $f'_{\sigma}(V_{ij})$ by $f'_{\sigma}$
and the designated basis of $f'_{\sigma}(V_{ij})$ is mapped to
the designated basis of $f'_{\pi}(f'_{\sigma}(V_{ij}))$ by $f'_{\pi}$.

Denote $W=f'_{\pi}(f'_{\sigma}(V_{ij}))$.
Then, $(f'_{\pi\sigma})^{-1}(W)$ is an irreducible invariant 
subspace and its designated basis is mapped to the 
designated basis of $W$ by $f'_{\pi\sigma}$. This is equivalent
to $(f'_{\pi\sigma})^{-1}$ mapping the designated basis of 
$W$ to the designated basis $(f'_{\pi\sigma})^{-1}(W)$.

By combining these two statements, we see that }
Therefore,
$(f'_{\pi\sigma})^{-1}f'_{\pi}f'_{\sigma}$ maps
the designated basis of $V_{ij}$ to $c$ times the designated basis 
of $(f'_{\pi\sigma})^{-1}f'_{\pi}f'_{\sigma}(V_{ij})=V_{ij}$.

It remains to show that $c$ is the same for all irreducible
subspaces $V_{ij}$ contained in $V_i$.
Let $c_j$ and $c_{j'}$ be the values of $c$ for $V_{ij}$ and
$V_{ij'}$.
Consider the subspace
\[ W=\{ x+h_{ijj'}(x) |x\in V_{ij} \} .\]
By Claim \ref{Form}, this is an irreducible invariant subspace. Now,
$(f'_{\pi\sigma})^{-1}f'_{\pi}f'_{\sigma}$ maps it to
\[ W'=\{ c_j x + c_{j'} h_{ijj'}(x) |x\in V_{ij} \} = \]
\[ \{ x + \frac{c_j}{c_{j'}}  h_{ijj'}(x) |x\in V_{ij} \} .\]
The invariance of $W$ means that $W'=W$ and $c_j=c_{j'}$.

Therefore, $c_j$ are all equal.
This means that $(f'_{\pi\sigma})^{-1}(x)=c_j x$ for all $x\in V_i$
because the designated bases of $V_{ij}$ together form a basis for
entire subspace $V_i$.

Unfortunately, arguments of this type (composing $f_{\pi}$ with
an appropriate transformation that fixes all $U_i$) cannot
be used to eliminate phases $c_{\pi, \sigma, i}$.

The reason for this is that there exist so-called {\em projective 
representations}. A projective representation is a set of maps $f_{\pi}$ 
such that $f_{\pi} f_{\sigma}=c_{\pi, \sigma} f_{\pi\sigma}$,
$c_{\pi, \sigma}\in\bbbc$. It is known that the symmetric group has
projective representations which are not equivalent to any
of the usual representations\cite{HH}.

One possible solution would be to use the standard forms of projective 
representations which are quite well studied\cite{HH}. 
However, to be able to use them,
we would need to show that the multiplicative constants
$c_{\pi, \sigma, i}$ are the same for all $V_i$ 
(or show that we can 
split all $V_i$ in several groups so that
$c_{\pi, \sigma, i}$ is
the same within one group) and we do not know if this is possible.

Our solution is to replace $f'_{\pi}$ by transformations
$f''_{\pi}$ on a larger space $V_1\otimes V_1^{*} + \ldots
V_k\otimes V_k^*$ so that $f''_{\pi}f''_{\sigma}=f''_{\pi\sigma}$.
Then, $f''_{\pi}$ form a representation in the usual sense and
we can analyse them similarly to section \ref{Simple}.

\subsection{Solving the problem with phases}

We split $V_1, \ldots, V_k$ into equivalence classes $\V_1, \ldots \V_l$.
$V_i$ and $V_j$ are in one class if there is a $\pi\in S_M$ such that
$f_{\pi}(V_i)=V_j$.
Let $W_i$ be the union of all $V_j$ that belong to $\V_i$.
Then, $f_{\pi}(W_i)=W_i$ for any $\pi\in S_M$
(because $f_{\pi}$ maps every $V_j\in \V_i$ to some $V_{j'}\in\V_i$).
Therefore, we can look at each $W_i$ separately.

\begin{lemma}
\label{bound-difference2}
Let $X, Y\in\hX$.
Then, for any $t$, 
\[ \dim X\cap W_t  - \dim X \cap Y\cap W_t
\leq \sqrt{\frac{4cn}{M}}\dim W_t .\]
\end{lemma}

\proof
To simplify the notation, assume that 
$W_t=V_1\oplus V_2 \ldots \oplus V_l$.

Consider the linear space 
$W'_t=V_1\otimes V^{*}_1 \oplus \ldots V_l\otimes V^{*}_l$
and the linear transformations $f''_{\pi}=f'_{\pi} \otimes f_{\pi}$.
These linear transformations form a representation because
\[ f'_{\pi\sigma}\otimes (f'_{\pi\sigma})^*=
c_{\pi, \sigma, i} f'_{\pi} f'_{\sigma} \otimes 
c^*_{\pi, \sigma, i} (f'_{\pi})^* (f'_{\sigma})^* = \]
\[ f'_{\pi} f'_{\sigma} \otimes (f'_{\pi})^* (f'_{\sigma})^* =
f''_{\pi} f''_{\sigma} \]
on every $V_i\otimes V^*_i$.

Let 
\[ X'= \oplus_{i=1}^l (X\cap V_i)\otimes (X\cap V_i)^{*} \]
be the subspace of $W_t^*$ corresponding to $X$.
Then, $f'_{\pi}(X)=Y$ implies $f''_{\pi}(X')=Y'$.
(To see this, consider one of $(X\cap V_i)\otimes (X\cap V_i)^{*}$.
Assume that $f'_{\pi}$ maps $V_i$ to $V_{i'}$.
Then, $f'_{\pi}(X)=Y$ implies
$f'_{\pi}(X\cap V_i)=Y\cap V_{i'}$ and
\[ f''_{\pi}((X\cap V_i) \otimes (X\cap V_i)^*)=
  (Y \cap V_{i'}) \otimes (Y\cap V_{i'})^{*} .\]
Combining these equalities for all $V_i$ gives 
$f''_{\pi}(X')=Y'$.)

In particular, $f''_{\pi}(X')=Y'$ means that
$X'$ is invariant under all $\pi\in S_M$
satisfying $\pi(X)=X$.
Therefore, by Lemma \ref{bound-difference}, 
\begin{equation}
\label{e1} 
\dim X'-\dim X' \cap Y'\leq \frac{4cn}{M} \dim W'_t .
\end{equation}
We use this inequality to derive a bound on 
$\dim X\cap W_t- \dim X\cap Y\cap W_t$. 
To do this, we relate the dimensions of 
$X\cap V_i$ and
$X' \cap (V_i \otimes V^*_i)$.
First, notice that we have
\begin{equation} 
\label{e2}
X\cap W_t=\oplus_{i=1}^l (X\cap V_i) 
\end{equation}
because $X$ is invariant under $G$ and, therefore, can
be written as a sum of irreducible invariant subspaces
(and each of these irreducibles is contained in some $V_i$).
The same is true about $Y$ and $X\cap Y$:
\begin{equation}
\label{e3} 
X\cap Y\cap W_t=\oplus_{i=1}^l (X\cap Y\cap V_i) 
\end{equation}
Let $d_i$ and $d'_i$ be the dimensions of $X\cap V_i$ and
$X\cap Y\cap V_i$. 
Then, (\ref{e2}) and (\ref{e3}) 
imply that $\dim X\cap W_t=\sum_{i=1}^l d_i$, 
$\dim X\cap Y\cap W_t=\sum_{i=1}^l d'_i$ and
\[\dim X\cap W_t-\dim X\cap Y\cap W_t=\sum_{i=1}^l 
(d_i-d'_i) .\]
If we look at $V_i\otimes V^*_i$, then 
\[ X' \cap (V_i\otimes V^*_i)= (X\cap V_i) \otimes (X\cap V_i)^*. \]
This implies $\dim X'\cap (V_i\otimes V^*_i)=d_i^2$ and
$\dim X'=\sum_i d_i^2$.
Similarly, $\dim X'\cap Y'=\sum_{i=1}^l {d'_i}^2$.

Let $d$ be the dimension of $V_1$. 
Then, the dimensions of $V_2$, $\ldots$, $V_l$ are $d$ as well
because, for every $i\in\{2, \ldots, l\}$,  there is a unitary $f_{\pi}$ 
such that $f_{\pi}(V_1)=V_i$.
Therefore, $\dim W_1=l d$.
Also, $\dim W'_t=l d^2$ because $\dim V_i\otimes V^*_i=d^2$ for
every $i\in\{1, \ldots, l\}$.
Hence, we have
\[ \frac{\dim X\cap W_t-\dim X\cap Y\cap W_t}{\dim W_t}= 
\frac{\sum_{i=1}^m (d_i-d_i)}{m d}=\]
\[\frac{1}{m}\sum_{i=1}^m \sqrt{\frac{(d_i-d'_i)^2}{d^2}} \leq 
 \frac{1}{m}\sum_{i=1}^m \sqrt{\frac{(d_i-d'_i)(d_i+d'_i)}{d^2}} =\]
\[ \frac{1}{m}\sum_{i=1}^m \sqrt{\frac{(d^2_i-d_i^{'2})}{d^2}} .\]
Convexity of the square root implies that this is at most
\[ \sqrt{\frac{\sum_{i=1}^m (d^2_i-d^{'2}_i)}{m d^2}}=
\sqrt{\frac{\dim X'-\dim X'\cap Y'}{\dim W'_t}} .\]
Equation \ref{e1} implies that this is at most
$\sqrt{(4cn)/M}$.
This completes the proof of lemma.
\rqed

With Lemma \ref{bound-difference2}, we can finish the proof
similarly to the simple case (section \ref{Simple}).
By summing over $W_t$'s, we get
\[ \dim X- \dim X \cap Y \leq \sqrt{\frac{2cn}{M}} \sum_{t} \dim
W_t =
 \sqrt{\frac{2cn}{M}} 2^n .\]
Therefore,
\[ \frac{\dim X\cap Y}{\dim X} 
\geq \frac{2^{n-k} - \sqrt{\frac{2cn}{M}} 2^n}{2^{n-k}} = 
1-\frac{2^{k+1}\sqrt{cn}}{\sqrt{M}} .\]
If $\frac{2^{k+1}\sqrt{cn}}{\sqrt{M}}\geq \frac{1}{\poly(m)}$,
then $m=(2k+\log n)(1+o(1))$.
This completes the proof of Theorem \ref{main-theorem}.
\rqed

\input epsf.sty

\section{Representation theory} \label{repthy}

\begin{description}
\item[Representation.] A representation $\rho$ of a group $G$ is a
  homomorphism $\rho$ from $G$ to the group of linear transformations $GL(V)$
  of a vector space $V$.
  This means that, for any $g,h \in G$, $\rho(gh)=\rho(g) \rho(h)$.  
  If the mapping $\rho$ is clear from the context, we often
  call the space $V$ itself representation of $G$.

\item[Irreducibility.] We say that a subspace $W$ is an
  \emph{invariant} subspace of a representation $\rho$ if 
  $\rho(g) W \subseteq W$
  for all $g \in G$. 
  In order for $W$ to be an invariant subspace for $\rho$, it must be
  simultanously fixed under all $\rho(g)$.
  The zero subspace and the subspace $V$ are always invariant.  If no
  nonzero proper subspaces are invariant, the representation is said
  to be \emph{irreducible}.

\item[Isomorphism.]
  Two representations $\rho:G \to GL(V)$ and $\rho':G\to GL(W)$ are
  isomorphic if there is a bijective linear map $\varphi:V\to W$ such that
  $\varphi \rho(g)=\rho'(g)\varphi$ for any $g\in G$.

\item[Schur's Lemma.]
  If $\rho$ and $\rho'$ are two irreducible representations
  and $\varphi$ is an isomorphism between them, then any other
  isomorphism $\varphi'$ between $\rho$ and $\rho'$ 
  is $c \varphi$ for some constant $c\in \bbbc$.

  Schur's lemma is usually stated for finite groups. 
  However, if the representation is unitary (as in this paper),
  it is also true for infinite groups.
 
\item[Decomposition.] When a representation \emph{does} have a nonzero
proper invariant subspace $V_1 \subset V$, it is always possible to
find a complementary subspace $V_2$ (so that $V = V_1 \oplus V_2$)
which is also invariant. Since $\rho(g)$ fixes $V_1$, we may let
$\rho_1(g)$ be the linear map on $V_1$ given by $\rho(g)$. It is not
hard to see that $\rho_1: G \to GL(V_1)$ is in fact a
representation. Similarly define $\rho_2(g)$ to be $\rho(g)$
restricted to $V_2$. Since $V = V_1 \oplus V_2$, the linear map
$\rho(g)$ is completely determined by $\rho_1(g)$ and $\rho_2(g)$, and
in this case we write $\rho = \rho_1 \oplus \rho_2$.  
  
\item[Complete Reducibility.] Repeating the process described above, a
  representation $\rho$ may be written 
  $\rho = \rho_1 \oplus \rho_2 \oplus \ldots \oplus
  \rho_k$, where each $\rho_i$ is irreducible. 

\item[Irreducible representations of $S_M$.]
  In this paper, we use representations of the symmetric 
  group $S_M$. The irreducible representations of $S_M$ may be placed 
  into one-to-one correspondence with the partitions of $n$. 
  A \emph{partition} of $M$ is a sequence $(\lambda_1, \ldots, \lambda_k)$ 
  of positive integers, with $\lambda_1 \geq
  \ldots \geq \lambda_k$ for which $\sum \lambda_i = M$.
  It is customary to identify
  the partition $\lambda = (\lambda_1, \ldots, \lambda_k)$ 
  with a diagram consisting of $k$
  rows of boxes, the $i$th row containing $\lambda_i$ boxes. 
  We will let $\lambda$
  stand for both the partition and the associated diagram. For example,
  the diagram corresponding to the partition 
  $\lambda = (4, 4, 2, 1)$ is shown
  in figure~\ref{fig1}.

\begin{figure}
\label{fig1}
\begin{center}
\epsfxsize=1.5in
\hspace{0in}
\epsfbox{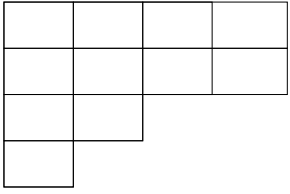}
\caption{\it The Young diagram of $\lambda=(4, 4, 2, 1)$.}
\end{center}
\end{figure}

  The irreducible representation associated with $\lambda$ is denoted
  $\rho_\lambda$. 
  There is an explicit formula for the dimension of $\rho_\lambda$.
  This involves the notion of a \emph{hook}: for a cell $(i,j)$ of a
  Young tableau $\lambda$, the $(i,j)$-hook $h_{i,j}$ is the collection of
  all cells of $\lambda$ which are beneath $(i,j)$ (but in the same column)
  or to the right of $(i,j)$ (but in the same row), including the cell
  $(i,j)$.  The \emph{length} of the hook $\ell(h)$ is the number of cells 
  appearing in the hook. With this notation, the dimension of $\rho_{\lambda}$
  may be expressed:
  \begin{equation}
  \label{hook}
    dim \rho_\lambda = \frac{n!}{\prod_{i,j} \ell(h_{i,j})},
  \end{equation}
  this product being taken over all hooks $h$ of $\lambda$.
  Figure \ref{fig2} shows the hook lengths for the partition 
  $\lambda = (4, 4, 3, 1)$. 
  Formula (\ref{hook}) implies that the dimension of corresponding 
  representation is 
  \[ \frac{11 !}{7\cdot 5\cdot 3\cdot 2\cdot 6\cdot 4\cdot 2\cdot 3}=
  1320 .\]
\begin{figure}
\label{fig2}
\begin{center}
\epsfxsize=1.5in
\hspace{0in}
\epsfbox{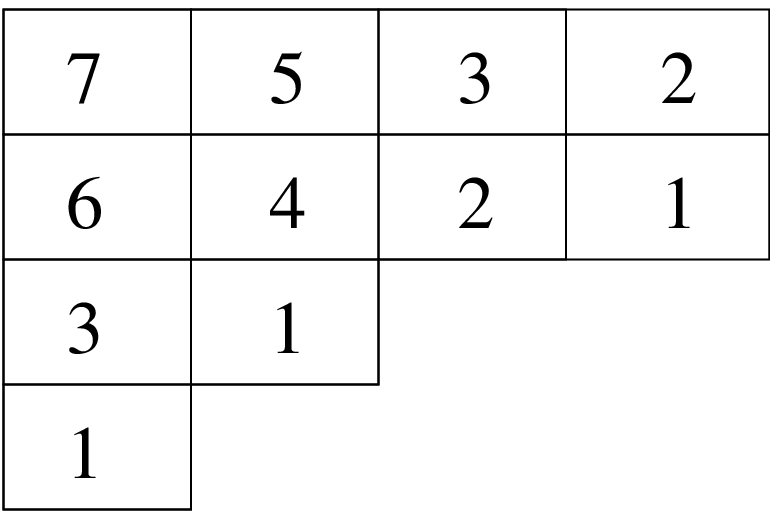}
\caption{\it The hook-lengths for (4, 4, 2, 1).}
\end{center}
\end{figure}

\item[Restriction.] A representation $\rho$ of a group $G$ is also
  automatically a representation of any subgroup $H$. 
  Note that even if a
  representation is irreducible over $G$, it may no longer be
  irreducible when restricted to $H$.

\item[Restriction from $S_M$ to $S_{M-1}$.]
  In particular, we will be considering the restrictions of irreducible 
  representations of $S_M$ to $S_{M-1}$. 
  Let $\lambda$ be a partition of $M$ and $\rho_\lambda$ be the
  corresponding irreducible representation. 
  Then, when we restrict to $S_{M-1}$, $\rho_{\lambda}$ decomposes
  into irreducible representations of $S_{M-1}$ in the following way:
  \[ \rho = \bigoplus_{\lambda^{-}} \rho_{\lambda^{-}} \]
  where ${\lambda^{-}}$ ranges over all shapes of size $M-1$ that can be
  obtained by deleting an ``inside corner'' from $\lambda$. (An inside
  corner is simply a point of the shape whose deletion leaves a legal
  shape.) 

  For example, the representation $\rho_{\lambda}$, $\lambda=(4, 4, 2, 1)$
  of $S_{11}$ decomposes into 3 irreducible representations of $S_{10}$.
  The Young diagrams of these representations are shown in Fig. \ref{fig3}. 
\begin{figure}
\label{fig3}
\begin{center}
\epsfxsize=2in
\hspace{0in}
\epsfbox{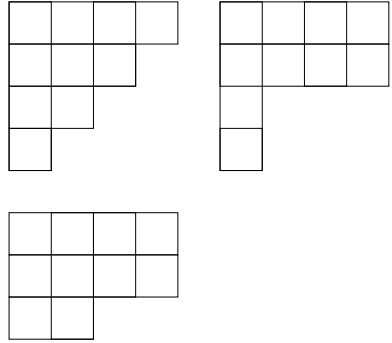}
\caption{\it The Young diagrams of representations of $S_{10}$ contained
in $\rho_{\lambda}$.}
\end{center}
\end{figure}
\end{description}

\end{document}